# An unbreakable cryptosystem

**Arindam Mitra**

Anushakti Abasan, Uttar Phalguni-7, 1/AF, Salt Lake,
Kolkata, West Bengal, 700064, India.

Abstract: The remarkably long-standing problem of cryptography is to generate completely secure key. It is widely believed that the task cannot be achieved within classical cryptography. However, there is no proof in support of this belief. We present an incredibly simple classical cryptosystem which can generate completely secure key.

Cryptography is an ancient art [1]. However, in the last century cryptography has emerged as a discipline of science. Now it has been working as a catalyst for the development of other disciplines ranging from number theory to quantum information to quantum computation.

Human being tried to develop unbreakable code from ancient times. In 1926, Vernam presented [2] a cryptosystem, known as one-time-pad system, which was later proved [3] to be a completely secure, that is, unbreakable system. However, one-time-pad system never became popular since key cannot be reused.

It has been pointed out [4-7] that the present computational strength can indirectly provide security in communication. Over the last four decades many computationally secure systems have been discovered [4-7]. Of them, some of the prototypes of DES and RSA [6,7] systems are widely regarded as safe and secure [4,5]. Although computationally secure systems are widely used, but many cryptographers and mathematicians have had some reservations about the very idea of computational security particularly where security is not even mathematically proven.

In 1983, in a seminal paper, Wiesner first pointed out [8] that quantum mechanics can ensure the security of data. Subsequently, Bennett and Brassard discovered [9] a quantum key distribution (QKD) protocol which can generate secure key with the help [10] of classical authentication scheme [11]. Since then many QKD protocols [12-15] have been designed. An alternative QKD system, which can itself generate secure key without any support of classical cryptosystem, has also been developed [16,17]. QKD system has already been implemented over several kilometers [13-15] and it has been proved to be completely secure [18-20] in some particular set-up. For public use, quantum cryptographic network is yet to be set up.

If quantum cryptographic network is set up all over the world and if that network is proved to be completely secure, still it would be advantageous to have a classical system which can generate completely secure key since our communication network is entirely classical. Apart from this advantage, it is always advantageous to have two routes to reach the destination. Next we shall present such classical cryptosystem.

System –I

Step 0#.
a). Two parties, called *A* and *B*, secretly share a key K which is a sequence of 2n random bits. The positions of 1s and 0s of K are denoted by r and p respectively.
b). They note down the random positions of 1s in the ascending order. It gives a sequence $K_0^r$ of positive integers. They note down the random positions of 0s in the ascending order. It gives another sequence $K_0^p$ of positive integers. This relationship between the sequences may be symbolically described as, $K \neg K^r$ and $K \neg K^p$.

Step 1#

a). A third party, called server, transmits a sequence $S_1$ of 2n random bits to both *A* and *B* through any communication channel.

b). *A* and *B* apply the pair of position keys $K^r$ and $K^p$ on $S_1$ and extract a pair of keys $k_1^r$ and $k_1^p$. It means bits of $k^r$ and $k^p$ are extracted from some random positions in $S_1$ determined by $K^r$ and $K^p$ respectively. That is, bits of $k^r$ are extracted from those positions in $S_1$ which are the positions of 1s in K and bits of $k^p$ are extracted from those bit positions in $S_1$ which are the positions of 0s in K. The extraction of keys may be symbolically described as $K^r: S_1 :: k_1^r$ and $K^p: S_1 :: k_1^p$.

Step 2#

a). The third party, called server, transmits a sequence $S_2$ of 2n random bits to both *A* and *B* through any communication channel.

b). *A* and *B* apply the same pair of position keys $K^r$ and $K^p$ on $S_2$, and extract a pair of keys $k_2^r$ and $k_2^p$. It means bits of $k^r$ and $k^p$ are extracted from some random positions in $S_2$ determined by $K^r$ and $K^p$ respectively. That is, bits of $k^r$ are extracted from those positions in $S_2$ which are the positions of 1s in K and bits of $k^p$ are extracted from those bit positions in $S_2$ which are the positions of 0s in K. The extraction of keys may be symbolically described as $K^r: S_2 :: k_2^r$ and $K^p: S_2 :: k_2^p$.

…………………………………………………………………………………………………..

Step i #

a). The third party, called server, transmits a sequence $S_i$ of 2n random bits to both *A* and *B* through any communication channel.

b). *A* and *B* apply the same pair of position keys $K^r$ and $K^p$ on $S_i$, and extract a pair of keys $k_i^r$ and $k_i^p$. It means bits of $k^r$ and $k^p$ are extracted from some random positions in $S_i$ determined by $K^r$ and $K^p$ respectively. That is, bits of $k^r$ are extracted from those positions in $S_i$ which are the positions of 1s in K and bits of $k^p$ are extracted from those bit positions in $S_i$ which are the positions of 0s in K. The extraction of keys may be symbolically described as $K^r: S_i :: k_i^r$ and $K^p: S_i :: k_i^p$.

**Illustration :**

Step 0#

a). Suppose the key K is

$$K = \{0, 1, 1, 1, 0, 1, 0, 1, 0, 0, 0, 1, 0, 1\}$$

b). The two positions keys are

$$K^r = \{2, 3, 4, 6, 8, 12, 14\}$$
$$K^p = \{1, 5, 7, 9, 10, 11, 13\}$$

Step 1 to i #

a). Suppose server transmits the following sequences to both $A$ and $B$ (none of the sequences is generated by any random number generator).

|   | 1 | 2 | 3 | 4 | 5 | 6 | 7 | 8 | 9 | 10 | 11 | 12 | 13 | 14 |
|---|---|---|---|---|---|---|---|---|---|----|----|----|----|----|
| $S_1 \equiv$ | 0 | 1 | 0 | 1 | 1 | 1 | 0 | 1 | 0 | 1 | 0 | 0 | 1 | 0 |
| $S_2 \equiv$ | 1 | 1 | 0 | 0 | 1 | 1 | 1 | 0 | 1 | 0 | 0 | 0 | 0 | 1 |
| $S_3 \equiv$ | 1 | 0 | 1 | 0 | 0 | 1 | 0 | 1 | 0 | 1 | 1 | 0 | 1 | 0 |
| $S_4 \equiv$ | 1 | 1 | 0 | 0 | 1 | 1 | 0 | 1 | 1 | 0 | 1 | 0 | 0 | 1 |
| . $\equiv$ | | | | | | | | | | | | | | |
| . $\equiv$ | | | | | | | | | | | | | | |
| . $\equiv$ | | | | | | | | | | | | | | |
| $S_i \equiv$ | 0 | 1 | 1 | 1 | 0 | 0 | 1 | 0 | 1 | 0 | 0 | 1 | 1 | 0 |

b). $A$ and $B$ extract the same pair of keys $k_i^r$ and $k_i^p$ applying the same pair of position keys $K^r$ and $K^p$ respectively on the same sequence $S_i$. $A$ and $B$ extract the following set of keys $R \in (k_1^r, k_2^r, .....k_i^r)$ by applying the position key $K^r$ on the sequences $S_1, S_2, .....S_i$ respectively.

$$
\begin{aligned}
K^r &\equiv 2\ 3\ 4\ 6\ 8\ 12\ 14 \\
k_1^r &\equiv 1\ 0\ 1\ 1\ 1\ 0\ 0 \\
k_2^r &\equiv 1\ 0\ 0\ 1\ 0\ 0\ 1 \\
k_3^r &\equiv 0\ 1\ 0\ 1\ 1\ 0\ 0 \\
k_4^r &\equiv 1\ 0\ 0\ 1\ 1\ 0\ 1 \\
. &\equiv\ .\ \ .\ \ .\ \ .\ \ .\ \ .\ \ . \\
. &\equiv\ .\ \ .\ \ .\ \ .\ \ .\ \ .\ \ . \\
. &\equiv\ .\ \ .\ \ .\ \ .\ \ .\ \ .\ \ . \\
k_i^r &\equiv 1\ 1\ 1\ 0\ 0\ 1\ 0
\end{aligned}
$$

Similarly, *A* and *B* extract the following set of keys $P \in (k_1^p, k_2^p, \ldots k_i^p)$ by applying the position key $K^p$ on the sequences $S_1, S_2, \ldots S_i$ respectively.

$$
\begin{aligned}
K^p &\equiv 1\ 5\ 7\ 9\ 10\ 11\ 13 \\
k_1^p &\equiv 0\ 1\ 0\ 0\ 1\ 0\ 1 \\
k_2^p &\equiv 1\ 1\ 1\ 1\ 0\ 0\ 0 \\
k_3^p &\equiv 1\ 0\ 0\ 0\ 1\ 1\ 1 \\
k_4^p &\equiv 1\ 1\ 0\ 1\ 0\ 1\ 0 \\
. &\equiv\ .\ \ .\ \ .\ \ .\ \ .\ \ .\ \ . \\
. &\equiv\ .\ \ .\ \ .\ \ .\ \ .\ \ .\ \ . \\
. &\equiv\ .\ \ .\ \ .\ \ .\ \ .\ \ .\ \ . \\
k_i^p &\equiv 0\ 0\ 1\ 1\ 0\ 0\ 1
\end{aligned}
$$

The extraction can be described as

$$K \neg K^r : (S_1, S_2 \ldots S_i) :: R \in (k_1^r, k_2^r, \ldots k_i^r)$$

$$K \neg K^p : (S_1, S_2 \ldots S_i) :: P \in (k_1^p, k_2^p, \ldots k_i^p)$$

Note that bit positions have been repeatedly used, not the bit values. Needless to say, Shannon's proof [2] does not disallow this indirect *reuse* of the shared key. Not only a pair/group of parties any number of pair/group of parties can extract their pair of keys from the same sequence, but none of the pair/group will know others' extracted key.

In cryptosystem, where establishing security against eavesdropping is the objective, key can be used for the following purposes: 1. To encrypt a meaningful message or even a key 2. To authenticate a person. 3. To generate a key. But it is still not known whether the same key can be used for all the three purposes.

We shall first consider that the generated key will be used in message encryption. We shall further assume that each extracted key will be separately used to encrypt each message. Each n-bit message M will be encrypted in n-bit key. The encryptions can be written as

$$C_1^r = k_1^r + M_1^r \qquad C_1^p = k_1^p + M_1^p$$

$$C_2^r = k_2^r + M_2^r \qquad C_2^p = k_2^p + M_2^p$$

$$\ldots\ldots\ldots\ldots \qquad \ldots\ldots\ldots\ldots$$

$$C_i^r = k_i^r + M_i^r, \qquad C_i^p = k_i^p + M_i^p$$

Suppose $A$ is the sender of message. Therefore, $A$ will transmit a set of cipher-texts $C^r \in (C_1^r, C_2^r \ldots C_i^r)$ to $B$ through any communication channel and $B$ will decode the set of messages $M^r \in (M_1^r, M_2^r \ldots M_i^r)$ applying the set of keys $R \in (k_1^r, k_2^r, \ldots k_i^r)$ respectively. $A$ can also transmit the set of cipher-texts $C^r \in (C_1^p, C_2^p, \ldots C_i^p)$ to $B$ through any communication channel and $B$ has to decode the set of messages $M^p \in (M_1^p, M_2^p \ldots, M_i^p)$ by applying the set of keys $P \in (k_1^p, k_2^p, \ldots k_i^p)$ respectively. The entire encryption process may be described as

$$K \neg K^r : (S_1, S_2, \ldots S_i) :: R \in (k_1^r, k_2^r, \ldots k_i^r) \mapsto (M_1^r, M_2^r \ldots M_i^r)$$

$$K \neg K^p : (S_1, S_2, \ldots S_i) :: P \in (k_1^p, k_2^p, \ldots k_i^p) \mapsto (M_1^p, M_2^p \ldots M_i^p)$$

In cryptography, a n-bit key/message is said to be completely secure if eavesdropper's probability of guessing the key is $p_g = \dfrac{1}{2^n}$. Let us prove that the system can deliver completely secure message.

*Proof:*

● We know, $K \neg K^r, K \neg K^p$

The key K is not known to eavesdropper. So, eavesdropper's probability of guessing $K^r$ or $K^p$ is

$$p_g = \frac{1}{2^n}$$

● We know, $K \neg K^r: S_i :: k_i^r$ and $K \neg K^p: S_i :: k_i^p$.

Applying $K^r$ on $S_i$ the key $k_i^r$ is extracted and applying $K^p$ on $S_i$ the key $k_i^p$ is extracted. So, eavesdropper's probability of guessing each extracted key is

$$p_g = \frac{1}{2^n}$$

● Note that,
$$C_i^r = k_i^r + M_i^r$$
$$C_i^p = k_i^p + M_i^p$$

● Eavesdropper gets $C_i^r$ and $C_i^p$. As $k_i^r$ and $k_i^p$ are used one time so, according to Shannon's proof [2], eavesdropper's probability of knowing $M_i^r$ and $M_i^p$ will be

$$p_g = \frac{1}{2^n}$$

Next we shall see that the system would crash if the two sets of data R and P are used as "password" or "PIN" for the purpose of authentication.

Eavesdropper is allowed to access the two sets of data R and P after their use. But from a single $k_i^r$ and single $k_i^p$ eavesdropper's probability of identifying a position-key is $\frac{1}{2^n}$. Suppose N keys $k_1^r, k_2^r, ....k_N^r$ are used as secure data for authentication. Therefore, after their use all the N keys are supposed to be available to eavesdropper. Eavesdropper gets the corresponding sequences $S_1, S_2, ... S_N$ from the communication channel. Now eavesdropper wants to correlate $k_1^r, k_2^r, \ldots k_N^r$ with $S_1, S_2, \ldots S_N$. This is possible if he can figure out the bit-position in the sequences $S_1, S_2, \ldots S_N$ from which the j-th bits of $k_1^r, k_2^r, \ldots k_N^r$ have been extracted. Suppose he observes that j-th bits of the keys $k_1^r, k_2^r, \ldots k_N^r$ are perfectly identical with i-th bits of the sequences $S_1, S_2, \ldots S_N$. But, due to accidental correlation this can happen with probability $\frac{1}{2^N}$. Therefore, the probability of wrongly identifying a bit position is $\frac{1}{2^N}$. So, the probability of correctly identifying a bit position is $p' = 1 - \frac{1}{2^N}$.

From the above analysis it can be concluded that if the same position key is used N times to extract N keys, and if all the N extracted keys become available to eavesdropper after their use, then eavesdropper's probability of correctly identifying n bit-positions, that is, a position-key, is

$$p = \left(1 - \frac{1}{2^N}\right)^n$$

Therefore, the system-I is unbreakable if extracted key is used in encryption, but it is quickly breakable, if extracted key is used as data for the purpose of authentication. Even unbreakable system will also crash if stealing of messages is allowed. By stealing N messages eavesdropper can decode N message-encrypting keys and then break the system with the same probability *p*.

The system-I can generate completely secure message. This is the strength of the system. The extracted keys are not independent keys to the eavesdropper. This is the weakness of the system. The above analysis suggests that a key should not be used more that one time, and *A* and *B* should use new new independent keys to extract independent keys. Without initially sharing many independent keys it is possible to exchange independent keys through our completely secure channel. That is, weakness of the system-I can be removed by its strength. The modified system is described below.

## System-II

Step 1#.

a). A third party server transmits a sequence $S_1$ of 2n random bits to both *A* and *B* through any communication channel.

b). *A* and *B* apply $K^r$ and $K^p$ of their shared key K on $S_1$ to extract a pair of keys $k_1^r$ and $k_1^p$ respectively. The extraction can be described as $K \neg K^r : S_1 :: k_1^r$ and $K \neg K^p : S_1 :: k_1^p$ c). *A* and *B* attach $k_1^r$ and $k_1^p$. The attached key can be denoted as $k_1 \equiv k_1^r .. k_1^p$. In the attached key, $k_1^r$ is placed before $k_1^p$.

d). *A* generates a 2n-bit independent key $X_1$ by random physical process and encrypts $X_1$ in $k_1$ and thereby prepares a cipher-text(key) $c_1 = k_1 + X_1$, which is transmitted to *B*.

e). *B* decodes $X_1$ from $c_1$ applying $k_1$.

f). A third party, called server, transmits another sequence $S_1^*$ of 2n random bits to both *A* and *B* through any communication channel.

g). *A* and *B* apply the same pair of position keys $X_1^r$ and $X_1^p$ of $X_1$ on $S_1^*$ to extract a pair of keys $x_1^r$ and $x_1^p$. This extraction may be described as $X_1 \neg X_1^r : S_1^* :: R_1^F \in (x_1^r)$ ; $X_1 \neg X_1^p : S_1^* :: P_1^F \in (x_1^p)$.

Step 2#.

a). A third party server transmits a sequence $S_2$ of 2n random bits to both *A* and *B* through any communication channel.

b). *A* and *B* apply $K^r$ and $K^p$ of their shared key K on $S_2$ to extract a pair of keys $k_2^r$ and $k_2^p$ respectively. The extraction can be described as $K \neg K^r : S_2 :: k_2^r$ and $K \neg K^p : S_2 :: k_2^p$.

c). They attach $k_2^r$ and $k_2^p$. The attached key is denoted as $k_2 \equiv k_2^r .. k_2^p$

d). *A* generates a 2n-bit key $X_2$ by random physical process and then encrypts $X_2$ in $k_2$ and thereby prepares a cipher-text(key) $c_2 = k_2 + X_2$, which is transmitted to *B*.

e). *B* decodes $X_2$ from $c_2$ applying $k_2$.

f). A third party, called server, transmits another sequence $S_2^*$ of 2n random bits to both *A* and *B* through any communication channel.

g). *A* and *B* apply the same pair of position keys $X_2^r$ and $X_2^p$ of $X_2$ on $S_2^*$ to extract a pair of keys $x_2^r$ and $x_2^p$. This extraction may be described as $X_2 \neg X_2^r : S_2^* :: R_2^F \in (x_2^r)$; $X_2 \neg X_1^p : S_2^* :: P_1^F \in (x_2^p)$ .

..................................................................................................................

Step i#.

a). A third party server transmits a sequence $S_i$ of 2n random bits to both *A* and *B* through any communication channel.

b). *A* and *B* apply $K^r$ and $K^p$ of their shared key K on $S_i$ to extract a pair of keys $k_i^r$ and $k_i^p$ respectively. The extraction can be described as $K \neg K^r : S_i :: k_i^r k_i^r$ and $K \neg K^p : S_i :: k_i^p$.

c). *A* and *B* attach $k_i^r$ and $k_i^p$. The attached key can be denoted as $k_i \equiv k_i^r .. k_i^p$

d). *A* generates a 2n-bit key $X_i$ by random physical process and encrypts $X_i$ in $k_i$ and thereby prepares a cipher-text(key) $c_i = k_i + X_i$, which is transmitted to *B*.

e). *B* decodes $X_i$ from $c_i$ applying $k_i$.

f). A third party, called server, transmits another sequence $S_i^*$ of 2n random bits to both *A* and *B* through any communication channel.

g). *A* and *B* apply the same position keys $X_i^r$ and $X_i^p$ of $X_i$ on $S_i^*$ to extract a pair of keys $x_i^r$ and $x_i^p$. This extraction may be described as $X_i \neg X_i^r : S_i^* :: R_i^F \in (x_i^r)$; $X_i \neg X_i^p : S_i^* :: P_i^F \in (x_i^p)$.

The following points may be pointed out.

1. The same position keys $X_i^r$ and $X_i^p$ are never reused to extract the final keys $x_i^r$ and $x_i^p$.

2. Exchanged key $X_i$ can neither be used as secure data nor in message encryption. If $X_i$ is used as secure data, then eavesdropper will know $k_i$ as $c_i = k_i + X_i$. If $X_i$ is used to encrypt a message $M_i$, then we have two equations $c_i = k_i + X_i$ and $C_i = X_i + M_i$. Here the same key $X_i$ is used two times to encrypt $k_i$ and $M_i$. So far complete security is concerned Shannon's proof [2] disallows it.

3. In each step, after the extraction of the pair of final keys $x_i^r$ and $x_i^p$, there is no need to store $X_i$ and $k_i$. They can be destroyed.

Let us prove that it is impossible for eavesdropper to figure out K and the pair of extracted sets $R \in (k_1^r, k_2^r, k_3^r \ldots k_N^r)$ and $P \in (k_1^p, k_2^p, k_3^p \ldots k_N^p)$. Only the final sets $R_1^F \in (x_1^r)$, $P_1^F \in (x_1^p)$, $R_2^F \in (x_2^r)$, $P_2^F \in (x_2^p)$, ......, $R_i^F \in (x_i^r)$, $P_i^F \in (x_i^p)$ are supposed to be available to eavesdropper after their use.

*Proof* :

● We know,

$K \neg K^r : (S_1, S_2 \ldots S_i) :: R \in (k_1^r, k_2^r, k_3^r \ldots k_i^r);\quad K \neg K^p : (S_1, S_2 \ldots S_i) :: P \in (k_1^p, k_2^p, \ldots k_i^p)$

$\Rightarrow (k_1^r .. k_1^p,\ k_2^r .. k_2^p, \ldots k_i^r .. k_i^p) \equiv U \in (k_1, k_2, \ldots, k_i,)$

To extract two sets R and P, the same pair of keys $K^r$ and $K^p$ are applied on $S_1, S_2 \ldots S_N$. But the extracted sets R and P are not used as secure data. Therefore, $U \in (k_1, k_2, \ldots, k_N)$ will not be available to eavesdropper after their use.

● The set of extracted keys $U \in (k_1, k_2, \ldots, k_N)$ are used in encryption.

$c_1 = k_1 + X_1;\ c_2 = k_2 + X_2; \ldots\ldots\ldots c_i = k_i + X_i$.

We have seen that transmitted keys $X_1, X_2, \ldots X_i$ are completely secure.

● Each exchanged key $X_i$ is used one time to extract a pair of keys.

$X_1 \neg X_1^r : S_1^* :: R_1^F \in (x_1^r);\qquad X_1 \neg X_1^p : S_1^* :: P_1^F \in (x_1^p)$

$X_2 \neg X_2^r : S_2^* :: R_2^F \in (x_2^r);\qquad X_2 \neg X_2^p : S_2^* :: P_2^F \in (x_2^p)$

............................................................................

$X_i \neg X_i^r : S_i^* :: R_i^F \in (x_i^r);\qquad X_i \neg X_i^p : S_i^* :: P_i^F \in (x_i^p)$

- We have seen that if the same position key is used N times to extract N keys, and if the extracted keys available to eavesdropper after their use, then eavesdropper's probability of identifying the position key is

$$p = \left(1 - \frac{1}{2^N}\right)^n$$

In this case, the same position key $X_i^r (X_i^p)$ is used one time to extract one key $x_i^r (x_i^p)$ respectively. So only one key $x_i^r (x_i^p)$ will be available to eavesdropper after its use. Therefore, in this case N =1. So eavesdropper's probability of knowing $X_i^r$ and $X_i^p$ from the extracted key $x_i^r$ and $x_i^p$ is $p = \frac{1}{2^n}$

- Given the final keys it is impossible for eavesdropper to figure out K. To eavesdropper, the final keys are independent keys.

The final sets $R_1^F \in (x_1^r)$, $P_1^F \in (x_1^p)$, $R_2^F \in (x_2^r), P_2^F \in (x_2^p)$, ……, $R_i^F \in (x_i^r)$, $P_i^F \in (x_i^p)$ can also be used for the following purposes. 1. To encrypt a meaningful message. After each step one of the parties can encrypt a n-bit message $M_i^r$ or $M_i^p$ in $x_i^r$ or $x_i^p$ to prepare cipher-text $C_i^r = x_i^r + M_i^r$ or $C_i^p = x_i^p + M_i^p$. 2. To authenticate a channel as secure data. 3. To generate a key.

The system-II can generate independent uncorrelated keys and the probability of breaking the system is $\frac{1}{2^n}$. To break the presented system, eavesdropper has no other option but to steal K. If eavesdropper steals K, he could totally break the system. Suppose eavesdropper fails to steal K. Still he can figure out K with probability less than 1, if he can steal more than one key $k_i$ or $X_i$ before their destruction.

Let us summarize the discussion. The system-I is unbreakable if the generated key is used only for message encryption. Here users need to share a key, but no random number generator is required. The system-II is unbreakable for all the three types of usage of the generated key - message encryption, authentication and key generation. Users need to share a key, but one of the users should possess a random number generator.

From the security point of view quantum and classical cryptosystem can be comparatively studied. In standard QKD system, two parties need to share a n-bit authentication key, and a random number generator is required to execute it. The probability of breaking completely secure [18-20] standard QKD system is $\frac{1}{2^n}$. To break it, eavesdropper has no other option but to initially steal the n-bit authentication key. We have seen that the same level of security can be achieved by the presented system. Of course both systems can peacefully coexist.

Apart from its use-value both standard and alternative QKD systems have theoretical importance. The strength of quantum cryptography is yet to be fully assessed. Recently, it has been observed [21,22] that quantum cryptography can provide complete security against cheating. Perhaps, herein lies the unparallel strength of quantum cryptography.

For the presented system, users need an open source of random bits and they should initially secretly share some random bits. However, sometimes open source may remain inaccessible and a common user may find it difficult to share random bits. We have found a prototype [23,24] of the presented system where, by secretly sharing some decimal numbers, parties can locally generate common key without using any communication channel.

*Email: mitra1in@yahoo.com